# Gaussian-shaped Optical Frequency Comb Generation For Microwave Photonic Filtering

Rui Wu, Christopher M. Long, Ehsan Hamidi, V.R. Supradeepa, Min Hyup Song, Daniel E. Leaird, *Senior Member, IEEE* and Andrew M. Weiner, *Fellow, IEEE*

***Abstract*—Using only electro-optic modulators, we generate a 41-line 10-GHz Gaussian-shaped optical frequency comb. We use this comb to demonstrate apodized microwave photonic filters with greater than 43-dB sidelobe suppression without the need for a pulse shaper.**

## I. Introduction

Phase modulated continuous-wave (CW) laser frequency combs have seen wide use in various applications such as wavelength division multiplexing (WDM) networks [1], optical arbitrary waveform generation (O-AWG) [2], and rapid arbitrary millimeter wave generation [3]. Using such optical frequency comb as a multiple carrier optical source offers new potential for achieving complex and tunable microwave photonic filters [4]. References [5] and [6] used a line-by-line pulse shaper to manipulate the amplitudes of individual spectral lines to obtain a carefully apodized Gaussian shape for achieving clean passband shapes with large sidelobe suppression, and also demonstrated a novel mechanism that allows tuning of the filter largely independent of passband shape. In reference [5], the initial comb has an irregular power spectrum. However reference [6] starts with our newly developed comb with record flatness and a large number of lines (38-lines within 1-dB variation out of a total 61 lines [7]), which eases pulse shaper requirements for implementation of high quality RF photonic filters. Although the generation of flat-topped optical frequency combs [8] [9] has drawn much attention, little work has been done to directly generate Gaussian-shaped optical frequency combs. In one report, Hisatake et al. demonstrated Gaussian-shaped comb generation based on spatial convolution of a slit and a periodically moving optical spot in a system based on electro-optic deflectors [10]. Here we introduce for the first time a directly generated 10-GHz Gaussian-shaped comb using only electro-optic modulators without the need for a pulse shaper. Therefore the implementation of RF photonic filters is much simplified, with lower optical loss and higher signal to noise ratio. In our current experiment, by employing a 41-line 10 GHz spaced directly generated Gaussian-shaped comb (37 lines closely matched to a Gaussian shape across a dynamic range of 20 dB), we implement filters with 43.2 dB sidelobe suppression. The results are improved over that reported in [5] and [6] and with a much more compact set-up.

This work was supported by the Naval Postgraduate School under grant N00244-09-1-0068 under the National Security Science and Engineering Faculty Fellowship program.

The authors are with the School of Electrical and Computer Engineering, Purdue University, West Lafayette, IN, 47907-2035, USA
(e-mail: rwu@purdue.edu; long25@purdue.edu; ehamidi@purdue.edu; supradeepa@purdue.edu;song55@purdue.edu;leaird@purdue.edu; amw@purdue.edu).

.

## II. Principle of Microwave Photonic Filters

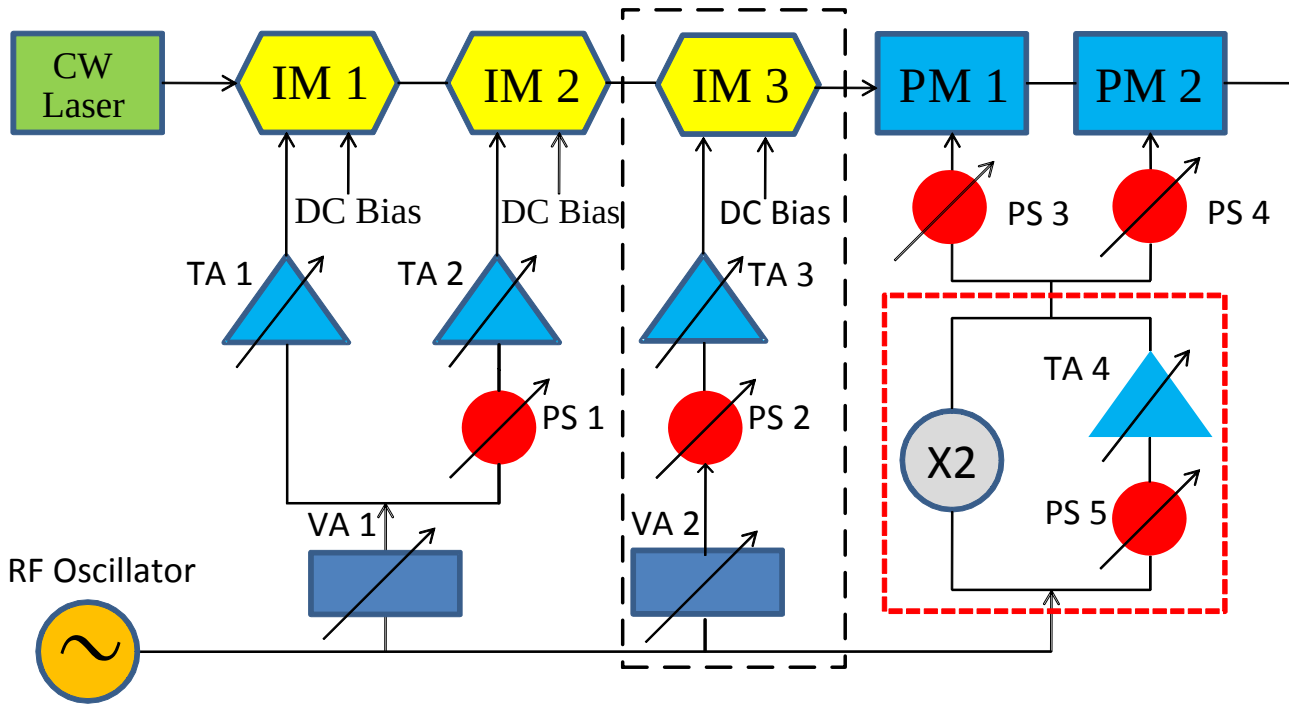

Fig. 1. Experimental setup for Gaussian-shaped comb generation. TA: tunable amplifier; VA: variable attenuator; PS: phase shifter; red dashed module: quasi-quadratic phase generation.

Fig. 1 shows the experimental setup of the 10-GHz Gaussian-shaped comb generator, which consists of three IMs and two PMs. We operate the RF oscillator at 10 GHz. At 10 GHz, the $V_\pi$ is ~9 V for the IM's and ~3 V for the PM's. The RF voltages delivered to IM1 and IM2 are both 0.5 $V_\pi$ zero to peak, and the RF voltage to IM3 is $V_\pi$. We cascade two phase modulators at their maximum RF input power (30 dBm) to double the total modulation index seen by the pulse and increase the number of comb lines.

The Gaussian-shaped comb generation mechanism is based on time-to-frequency mapping theory, where quadratic and periodic temporal phase causes the spectral envelope to mimic the input intensity profile to the phase modulators [11] [12] [13]. So for generating a Gaussian-shaped comb, the following two requirements should be met:

(i) Apply a quadratic phase. Applying a purely quadratic, periodic temporal phase is difficult, so here we apply a "quasi-quadratic" phase by combining the first and second harmonic of the sinusoidal drive signal with a power ratio of 24-dB and phase shift of 180° selected to suppress the $4^{th}$ order term of the cosine expansion of the phase - refer to the red dashed box in Fig. 1. This substantially improves the approximation to the target quadratic phase profile [7].

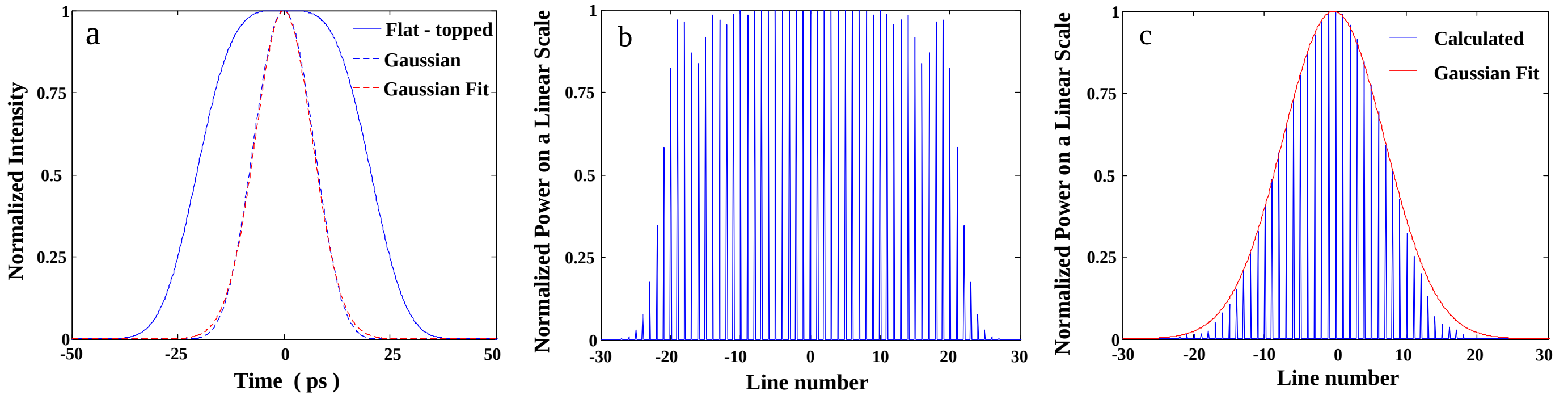

Fig. 2. Simulation results: (a) Solid blue: flat-topped pulse expected from the series combination of IM1 and IM2; Dashed blue: Gaussian pulse expected from the series combination of IM1, IM2 and IM3; Dashed red: Gaussian fit; (b) Simulated flat-topped comb spectrum (c) Simulated Gaussian-shaped comb spectrum.

(ii) Generate a Gaussian-shaped pulse. IM1 and IM2 are both biased at 0.5 $V_\pi$ with RF drive amplitude 0.5 $V_\pi$. IM3 is biased at 0 (maximum transmission) with RF drive amplitude $V_\pi$. We can get a flat-topped pulse with a series combination of IM1 and IM2 (without the black module in Fig. 1) and a very close approximation to a Gaussian pulse with the series combination of all three IMs as shown in Fig. 2(a). After applying the quasi-quadratic phase as described in (i), modeling indicates it is possible to obtain flat-topped or Gaussian-shaped combs as shown in Fig. 2 (b)-(c).

## III. PRINCIPLE OF MICROWAVE PHOTONIC FILTERS

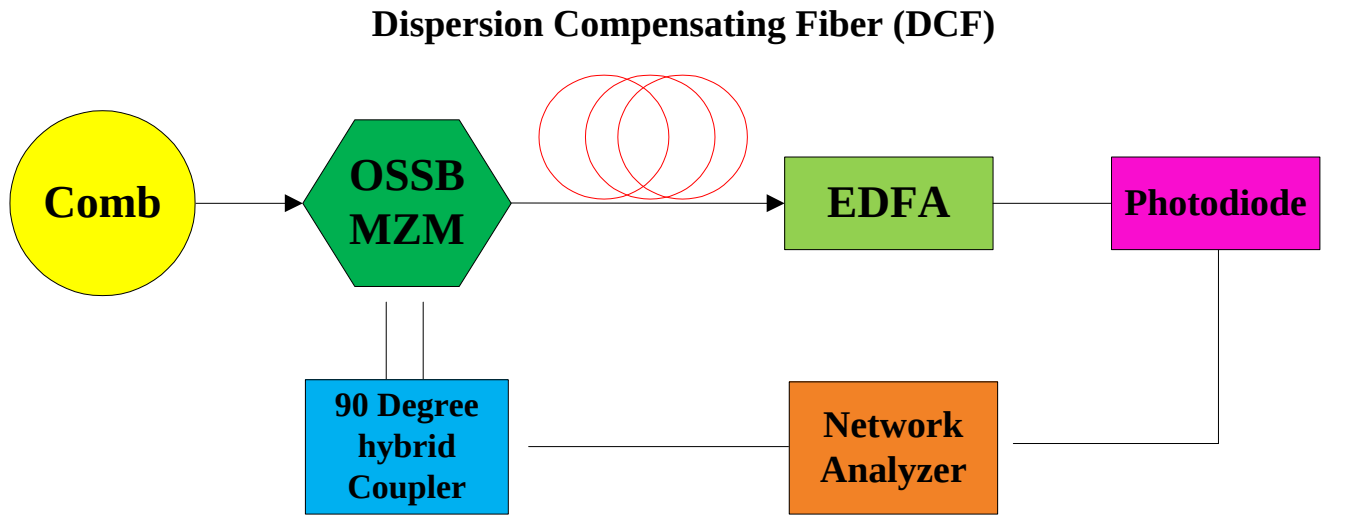

Fig. 3. Experimental setup of the microwave photonic filter based on Gaussian-shaped frequency comb. OSSB MZM: Optical Single-sideband Mach-Zehnder Modulator.

Without the use of a line-by-line pulse shaper, we employ the directly generated Gaussian-shaped comb as a multiple carrier optical source for RF photonic filtering and shape the filter pass bands.

Fig. 3 shows our RF photonic filtering setup. The comb is single-sideband modulated in a dual drive Mach-Zehnder modulator biased at the quadrature point and driven by a pair of RF signals with a 90 degree phase difference. The modulator output is sent through a dispersion compensating fiber (DCF) with specified dispersion of -1259.54 ps/nm at 1550 nm and relative dispersion slope (the ratio of dispersion slope to dispersion at 1550 nm) of 0.00455/nm, resulting in 96-ps relative tap delay between adjacent 10 GHz comb lines. The optical output signal, after being amplified by an EDFA is detected by a 22 GHz bandwidth photodiode and measured by a network analyzer over a span of 300 kHz-20GHz. The filter transfer function can be written as:

$$H(\omega_{RF}) \propto \sum_{n=0}^{n=N-1} e_n^2 e^{jnD2\pi\Delta f\,\omega_{RF}} \qquad (1)$$

where N is the total number of the taps, $e_n^2$ is the optical intensity of the $n^{th}$ tap, D is the fiber dispersion, $\Delta f$ is the repetition frequency of the optical comb, 10 GHz in our experiment, $D2\pi\Delta f$ is the tap delay between two adjacent taps, 96-ps in our experiment. The free spectral range (FSR) is the inverse of the tap delay, which can be tuned by changing the length of DCF.

## IV. EXPERIMENTAL RESULTS

Fig. 4 shows the experimental results. First, we turn off TA3 (in Fig. 1). Fig. 4(a) shows a 50-line 10-GHz flat-topped optical frequency comb spectrum measured at the photodiode (in Fig. 3). After turning on TA3, Fig. 4(b) shows a 41-line 10-GHz Gaussian-shaped optical frequency comb spectrum measured at the photodiode; the EDFA is adjusted to set the average power at the photodiode to 4.3 dBm, as used with the flat-topped comb. The standard deviation of the comb line amplitudes from a best-fit Gaussian profile is 0.42 dB for the central 37 lines and 0.26 dB for the central 35 lines, with an excellent match over 37 spectral lines across a 20-dB dynamic range. After applying the comb to our filter setup, we measured the filter transfer function with a network analyzer. We compare our experimental results with simulations that include an RF calibration factor which accounts for cable loss, modulator transfer function and the photodiode frequency response (high frequency roll-off). The RF calibration data can be obtained by measuring the link frequency response without the dispersion compensating fiber [5]. Fig. 4(c) shows the measured (blue) and simulated (red, after calibration) filter transfer functions using the flat-topped comb. At baseband, the filter has a 3-dB bandwidth of 100 MHz, with 18.1 dB sidelobe suppression. The passband at 10.4 GHz has a 3-dB bandwidth of 230 MHz with 14.4 dB sidelobe suppression. Modeling (based on the measured optical power spectrum) indicates the filter has a 3-dB bandwidth of 115 MHz and 230 MHz with 17.64 dB and 15.43 dB sidelobe suppression at the baseband and passband respectively, in very close agreements with our experiment. Fig. 4(d) shows the measured (blue) and simulated (red, after calibration) filter transfer functions using the above Gaussian-shaped comb. At the baseband, the filter has a 3-dB bandwidth of 120 MHz, with 43.2 dB sidelobe suppression.

The 10.4 GHz passband has a 3-dB bandwidth of 330 MHz with 35.3 dB sidelobe suppression. Modeling indicates the filter has a 3-dB bandwidth of 165 MHz and 330 MHz with 41.1 dB and 34.9 dB sidelobe suppression at baseband and the passband respectively, in very close agreements with our experiment. The filter FSR is determined by the length of DCF fiber. In our experiment, the filter FSRs with both comb shapes are 10.4 GHz in accordance with the 96-ps tap delays. By adding or decreasing the length of DCF we can change FSR, red-shifting or blue-shifting the filter pass band frequency accordingly.

## V. Conclusion

For the first time, we introduce a 10-GHz Gaussian-shaped optical frequency comb generation (37 lines across a 20 dB dynamic range out of a total 41 lines) using only electro-optic modulators. Based on this directly generated Gaussian-shaped optical frequency comb, we demonstrate 43.2 dB high sidelobe suppression microwave photonic filters. Because a line-by-line pulse shaper is no longer required, the experimental implementation is greatly simplified. The filter FSR can be tuned by changing the length of DCF. Also our directly generated Gaussian-shaped comb is fully compatible with the novel tuning approach, based on varying optical delay in an interferometer structure, demonstrated in [5] [6].

## Acknowledgment

We would like to thank Dr. Victor Torres-Company and Dr. Li Xu for valuable discussions.

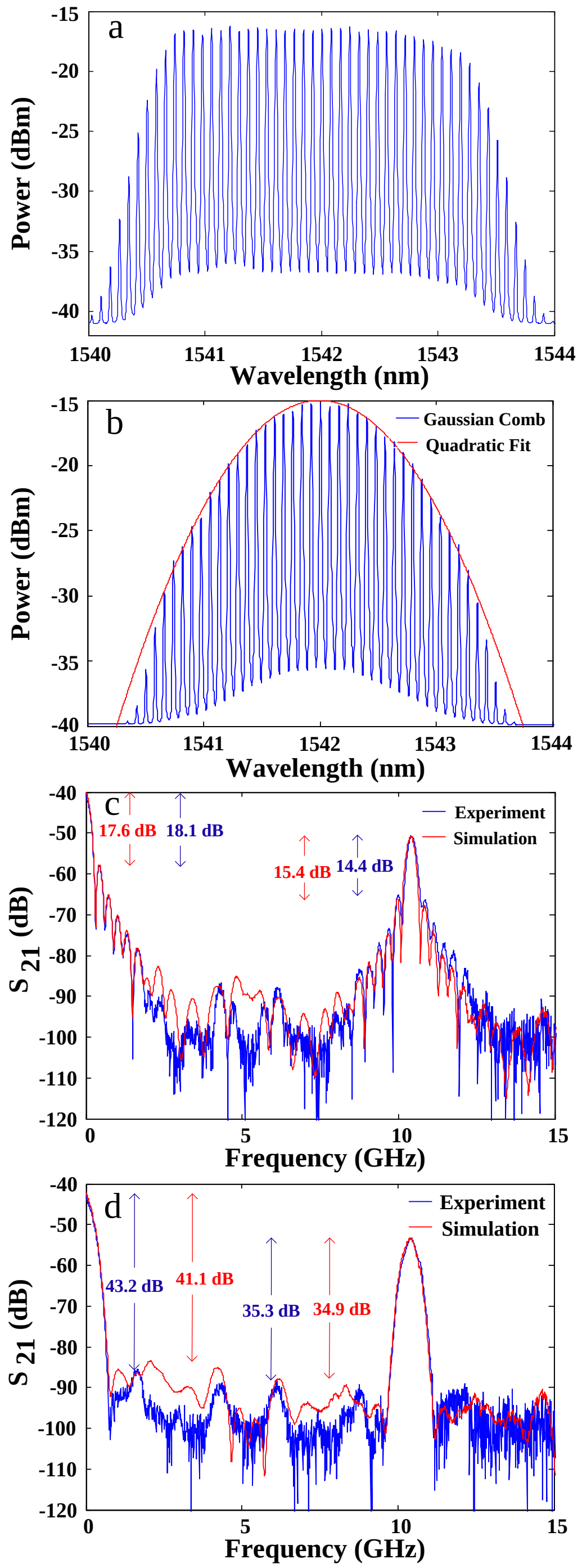

Fig. 4. Experimental results: (a) Flat-topped comb on a log scale; (b) Gaussian-shaped comb on a log scale; (c) Measured (blue) and simulated (red, after calibration) RF photonic filter transfer functions with flat-topped comb in (a); (d) Measured (blue) and simulated (red, after calibration) RF photonic filter functions with Gaussian-shaped comb in (b).